# Spin-dependent localization of helical edge states in a non-Hermitian phononic crystal

Junpeng Wu[1], Riyi Zheng[1], Jialuo Liang[1], Manzhu Ke[2], Jiuyang Lu[1,2*], Weiyin Deng[1,2*], Xueqin Huang[1*], Zhengyou Liu[2,3*]

[1]School of Physics and Optoelectronics and State Key Laboratory Luminescent Materials and Devices, South China University of Technology, Guangzhou 510640, China

[2]Key Laboratory of Artificial Micro- and Nanostructures of Ministry of Education and School of Physics and Technology, Wuhan University, Wuhan 430072, China

[3]Institute for Advanced Studies, Wuhan University, Wuhan 430072, China

*Corresponding author.
jylu@whu.edu.cn; dengwy@whu.edu.cn; phxqhuang@scut.edu.cn; zyliu@whu.edu.cn

**Abstract**

As a distinctive feature unique to non-Hermitian systems, non-Hermitian skin effect displays fruitful exotic phenomena in one or higher dimensions, especially when conventional topological phases are involved. Among them, hybrid skin-topological effect is theoretically proposed recently, which exhibits anomalous localization of topological boundary states at lower-dimensional boundaries accompanied by extended bulk states. Here we experimentally realize the hybrid skin-topological effect in a non-Hermitian phononic crystal. The phononic crystal, before tuning to be non-Hermitian, is an ideal acoustic realization of the Kane-Mele model, which hosts gapless helical edge states at the boundaries. By introducing a staggered distribution of loss, the spin-dependent edge modes pile up to opposite corners, leading to a direct observation of the spin-dependent hybrid skin-topological effect. Our work highlights the interplay between topology and non-Hermiticity and opens new routes to non-Hermitian wave manipulations.

Hermiticity plays an important role in quantum physics, while novel phenomena due to non-Hermiticity have attracted considerable interest recently [1–3]. Non-Hermitian systems are intrinsically distinct from the Hermitian ones since their spectra are in general complex and thus form delicate patterns [3–5]. One representative consequence is the widely studied non-Hermitian skin effect due to non-zero spectral winding or area, which enforces the bulk modes to accumulate towards the boundaries or surfaces of finite systems [6–12]. Moreover, higher-dimensional non-Hermitian systems are further enriched and can conceive phenomena such as higher-order skin effects [13–19]. Recently the interplay between conventional topological effects and the non-Hermitian skin effect has received significant attention [20–27]. Hybrid skin-topological effect (HSE), allowing topological chiral edge states to accumulate towards the corners of a finite-sized sample, has recently been proposed [22–25] and observed in electric circuit [26] and photonic crystal [27]. However, the bridge between non-Hermitian skin effect and helical edge states remains undiscovered.

Phononic crystals (PCs) over the last few years show great potential in practical realizations of topologically nontrivial phases [28–30]. One of the early attempts is the acoustic Chern insulator where active devices were presented to break time-reversal symmetry [31–35]. By inspecting the construction of internal degrees of freedom in PCs, time-reversal-invariant acoustic topological insulators such as acoustic spin-Chern insulators and acoustic valley Hall insulators were proposed and experimentally confirmed [36–43]. The recent past has witnessed the emergence of non-Hermitian acoustics [44–53]. The introduction of non-Hermiticity in acoustic systems can be implemented with the help of external active circuits [46]. While a considerable amount of research has focused on topological and non-Hermitian aspects of PCs, a comprehensive investigation of these two properties is rare. As such, exploring the non-Hermitian topological effects, such as the gain-loss-induced HSE, is in demand. A more feasible and stable approach is to adopt only absorbing materials in observing HSE and maintain the system passive [47,48].

Here we experimentally demonstrate the spin-dependent HSE (SHSE) in a non-Hermitian PC with a staggered loss distribution, as a correspondence of the gain-loss-induced HSE in a spinful system. Our design is based on a bilayer structure and interlayer couplings contribute to the layer pseudospin-orbit interactions [39]. Before losses are introduced, the elaborately designed PC is an acoustic analogue of the Kane-Mele model, providing gapless helical edge states at the boundaries corresponding to opposite spins. By introducing the staggered losses, the helical edge states are observed to accumulate at different corners depending on their spins, demonstrating the SHSE for acoustic waves. Simulated and experimental results consistently verify the nontrivial topological phenomena with and without non-hermiticity.

We start from a typical lattice of the Haldane model as shown in the left panel of Fig. 1(a), where $t_1$ and $t_2$ denote the strengths of the nearest and next nearest neighborhood hoppings. Non-Hermiticity is further introduced by imposing gain and loss in the inequivalent sites A and B (with strength $\gamma$). The nontrivial flux in the Haldane model requires time-reversal symmetry breaking, which set difficulty in realization. An alternative practical proposal is to extend to a bilayer lattice [54,55] as shown in the right panel Fig. 1(a): the nearest hoppings are simply duplicated in each layer, while the next nearest hoppings are replaced by a pair of opposite hoppings connecting different layers. Accordingly, the unit cell boxed by dashed green lines in Fig. 1(a) is modulated to that shown in Fig. 1(b). The related Hamiltonian can thus be written as

$$H = t_1 \left[\cos\frac{k_y}{\sqrt{3}} + 2\cos\frac{k_x}{2}\cos\frac{k_y}{2\sqrt{3}}\right]\sigma_x + t_1\left[\sin\frac{k_y}{\sqrt{3}} - 2\cos\frac{k_x}{2}\sin\frac{k_y}{2\sqrt{3}}\right]\sigma_y - 2t_2\left[\sin k_x - 2\sin\frac{k_x}{2}\cos\frac{\sqrt{3}k_y}{2}\right]\sigma_z\tau_y - i\gamma\sigma_z, \qquad (1)$$

where $\sigma$ and $\tau$ are Pauli matrices denoting the hoppings between sites A and B in the same layer and those between different layers (acting as layer pseudospins or spins for simplicity), and the lattice constant is set to 1 for simplicity. The Hermitian part of $H$ in Eq. (1) (the first three terms) is just the Kane-Mele model in the absence of Rashba spin-orbit coupling and on-site energies [56,57], which commutes with the layer pseudospin operator $\tau_y$. Therefore, the Hermitian part of $H$ supports a pair of precisely degenerate band structures that represent opposite spins as shown in Fig. 1(c). Note that the non-Hermitian part of $H$ [the last term in Eq. (1)] is also independent on the layer pseudospin. In this regard, the whole Hamiltonian $H$ can be viewed as a direct sum of a non-Hermitian Haldane model [as shown in Fig. 1(a)] and its time-reversal counterpart with opposite $t_2$, and the results in the non-Hermitian Haldane model [24,25] can be straightforwardly generalized in this bilayer model.

By introducing the gain and loss non-Hermiticity in the Kane-Mele model, hybrid skin-topological states can occur for different spins, that is the helical topological edge modes that originally propagate along the open boundaries are localized at different corners in a finite lattice. Specifically, in our bilayer model of different layer pseudospins, the spin-up polarized edge states are found to be localized at the upper left corner [Fig. 1(d)] whereas the spin-down polarized ones are localized at the lower right corner [Fig. 1(e)], revealing the SHSE. In these figures, the field distributions are obtained by summing up the squared magnitudes of all the edge state fields of opposite spin polarizations, respectively. To further explore the underlying physics of the emergence of SHSE, we can view the two corners as the domain walls separating the zigzag edges of gain and loss [58]. Considering the one-dimensional boundary of the sample, e.g., a closed path I → II → III → IV → I in Figs. 1(d) and 1(e), one obtains from an effective Hamiltonian [59] that the spin-dependent helical edge states of eigenenergy $E$ have the form

$$\psi_\pm(x) = \frac{N}{\sqrt{2}}\begin{pmatrix}1\\\pm i\end{pmatrix}\exp\left[i(\pi \pm \operatorname{Re} E/v_{\text{eff}})x \pm v_{\text{eff}}^{-1}\int_0^x (\gamma(x') - \operatorname{Im} E)dx'\right], \quad (2)$$

where $\pm$ represents the spin-up or -down component, $N$ is a normalization factor, $v_{\text{eff}}$ stands for the speed of helical edge states, and $x$ runs over all the open boundaries along the closed path from 0 to $L$ (length of the closed path). In the integrand, $\gamma(x)$ describes the gain/loss distribution along the boundary, which takes the form $\gamma_{\text{eff}}[\operatorname{sgn}(x - L/4) + \operatorname{sgn}(3L/4 - x) - 1]/2$ with sgn representing the sign function. Both $v_{\text{eff}}$ and $\gamma_{\text{eff}}$ are extracted from the calculated projected band structures of the helical edge states. Note that the finite closed path sets constraints for $E$ as $\operatorname{Im} E = L^{-1}\int_0^L \gamma(x)dx$ and $\pi \pm \operatorname{Re} E/v_{\text{eff}} = 2\pi n/L$ with $n$ being integers. As a result, $\operatorname{Im} E$ vanishes with the above $\gamma(x)$, yielding a real eigenenergy for each helical edge states. The distribution of $\gamma(x)$ also offers a natural way to distinguish two types of domain walls: one domain wall resides at the rising edge of $\gamma(x)$, abruptly from low to high values, while the other at the falling edge. Moreover, the integral in Eq. (2) takes the form $\pm|\gamma_{\text{eff}}\Delta x/v_{\text{eff}}|$ near those domain walls with $\Delta x$ being the deviation distance from the domain wall, which means that the helical edge states with opposite spins share the same localization length of $|v_{\text{eff}}/\gamma_{\text{eff}}|$. As shown in Fig. 1(f), the spin-dependent localizations of $\psi_\pm$ agree well with the numerical results of the tight-binding model, announcing the effectiveness of Eq. (2) in capturing the SHSE. The SHSE can also been viewed as the result of the non-zero windings of the spectra loops of the topological edge states [59].

To experimentally verify the existence of the SHSE, we start by mapping the Hermitian part of $H$ into a PC for acoustic waves. The PC sample is fabricated by 3D printing with its photo shown in Fig. 2(a). The unit cell of the PC consists of four connected hexagonal prism cavities as illustrated Fig. 2(b), where only the air-filled region is shown. The lattice constant is $a = 6.0$ cm, and the height and side length of the hexagonal prism cavity are $h_c = 0.50a$ and $l_c = 0.19a$, respectively. Here we focus on acoustic dipole modes in the hexagonal prism cavities, and both positive and negative couplings can be achieved with elaborately designed waveguide connections between cavities [60,61]. In experiment, the acoustic pressure fields are captured by a microphone positioned on the central axis of each cavity, approximately 5 mm from its outward-facing base surface. By Fourier transforming the scanned acoustic pressure fields, the measured bulk band structures are obtained [Fig. 2(c)], agreeing well with the simulated ones. The bulk bands are topologically nontrivial with the topology characterized by the $\mathbb{Z}_2$ invariant and is equivalently reflected in the evolution of the Wannier centers. Figure 2(d) shows the Wannier center evolutions of the first two bands (of dipole modes) along the reciprocal lattice vector $\boldsymbol{b}_1$, exhibiting the winding numbers of $\pm 1$ ($\mathbb{Z}_2 = 1$). According to the bulk-boundary correspondence, a pair of gapless helical edge states emerge at the boundaries of the PC, as shown by the simulated dispersions of a zigzag-terminated ribbon [white curves in Figs. 2(e) and 2(f)]. These two helical edge states possess opposite spins, and by recombing the measured fields of the top and bottom cavities as $p_\pm = p_1 \pm ip_2$, the measured dispersions [colormaps in Figs. 2(e) and 2(f)] manifest that the spin-up and -down states have positive and negative group velocities, respectively.

Based on the validated acoustic realization of the Kane-Mele model, the mapping of the non-Hermitian part of $H$ is achieved by exposing the air-filled cavities of sublattice A (blue) to absorbing materials while leaving those of sublattice B (red) blank, resulting in a staggered distribution of on-site dissipations. Although the tight-binding model involves both gain and loss, only losses are introduced in PC system implementations; the ignorance of gain relocates the zero-energy point in the tight-binding model and has no impact on the emergence of the SHSE. In simulations, the losses are reflected by the imaginary parts of the sound speed, which is a function of spatial position with a staggered distribution. Single-cavity measurements show that the absorbing materials give rise to an imaginary part of about 8 m/s to the sound speed, in contrast to the intrinsic loss of about 3 m/s [59]. The examination of the non-Hermitian PC takes place on the diamond-shaped sample similar to those in Figs. 1(d)-1(e) as depicted in Fig. 3(a). Figure 3(b) shows the simulated complex eigen spectrum of the PC sample, while our focus lies on the eigenmodes that reside in the bulk gap (the shaded region). To quantify both the edge and corner localizations of each eigenmode $p^{(n)}$, the localization measure $L^{(n)}$ is defined as the product of a measure of the edge localization and a measure of the localization at two corners of interest [green shadings in Fig. 3(a)], i.e., $L^{(n)} = \Sigma'|p^{(n)}|^2/\Sigma|p^{(n)}|^2 \cdot \Sigma''|p^{(n)}|^2/\Sigma|p^{(n)}|^2$, where $\Sigma'$ and $\Sigma''$ sum up cavities on the boundaries and at the corners of interest while $\Sigma$ sums over all cavities. We plot $L^{(n)}$ for each eigenmode in the complex eigen spectrum, evidencing the expected localization of the eigenmodes within the bulk gap. The spin dependency of the localization of these eigenmodes characterized by $L^{(n)}$ can be unwrapped via the $p_\pm$ recombination. As expected from the underpinning tight-binding model, the spin-up polarized acoustic eigenmodes are localized at the upper-left corner of the sample while the spin-down polarized counterparts are localized at the lower-right corner, as shown in Figs. 3(c) and 3(d), respectively.

To verify the SHSE in the non-Hermitian PC in experiment, we set the measurement configuration as shown in Fig. 3(a), where two speakers are severally placed at the sample corners I and III for excitation and the responses are obtained by summing up the excited fields [59]. In this scenario, one would expect the field localizations to be inequivalent for certain spin polarizations at the corners II and IV due to the SHSE. The spin-polarized response fields $P_\pm$ is extracted by collecting $|p_\pm|^2$ within the bulk gap frequency range and from both excitations, where $|p_\pm|^2$ is the squared norm of the spin-up or -down polarized response normalized with respect to the maximum value. Figure 3(e) shows the spatial distributions of $P_\pm$ in the non-Hermitian PC, exhibiting inequivalent responses at one corner over the other. This observation is consistent with the corner localization pattern of the simulated spin-polarized eigenmodes, and directly demonstrates the unignorable participation of spin hybrid skin-topological modes in the response fields of the non-Hermitian PC. As a sharp comparison, the response fields $P_\pm(x)$ of the Hermitian PC, as illustrated in Fig. 3(f), are equivalent at the corners II and IV for each spin, indicating the absence of spin-dependent corner localization.

We further measure the spin-dependent transmission spectra at the corners II and IV. The transmission spectra are measured by gathering $|p_\pm|^2$ from the corner cavities [within the green shadings in Fig. 3(a)] and involving both excitations. Required by the SHSE, the transmission spectra for the non-Hermitian PC [Figs. 4(a) and 4(b)] show apparent inequivalence within the bulk gap frequency range (gray regions). Specifically, the spin-up states tend to accumulate at the corner IV, while the spin-down states prefer to localize at the corner II. Here the measurements (hollow symbols) and simulations (solid curves) show great consistency in the bulk gap. Again, in contrast, the transmission spectra for the Hermitian PC do not show significant differences for the spin-up or spin-down states, as depicted in Figs. 4(c) and 4(d).

In summary, the SHSE is experimentally verified in this work, which, before adding non-Hermiticity, is an ideal acoustic realization of the Kane-Mele model that hosts gapless helical edge states at PC boundaries. The non-Hermitian scheme we employ here includes only pure losses, offering capability of low cost and high stability in control. As such, we have observed the helical edge states pile up to specific corners depending on their spin features. Actually, the proposed lattice model and PC sample are not limited to the diamond shape for observation of SHSE, other sample shapes are investigated and by identifying the non-Hermitian distribution of domain walls one can find out the localization corner for the states of specific spin [59]. Our work exemplifies the richness in the interplay between topology and non-Hermiticity, and sheds light on observing non-Hermitian topological properties in acoustic wave systems. Moreover, the SHSE can also be extended to other systems, such as photonic crystal fibers [62–64], elastic wave systems [65–67], and electric circuits [26,54,68].

**Acknowledgements**

This work is supported by the National Key R&D Program of China (Nos. 2022YFA1404500, 2022YFA1404900), National Natural Science Foundation of China (Nos. 12074128, 12222405, 12374409), and Guangdong Basic and Applied Basic Research Foundation (Nos. 2021B1515020086, 2022B1515020102).

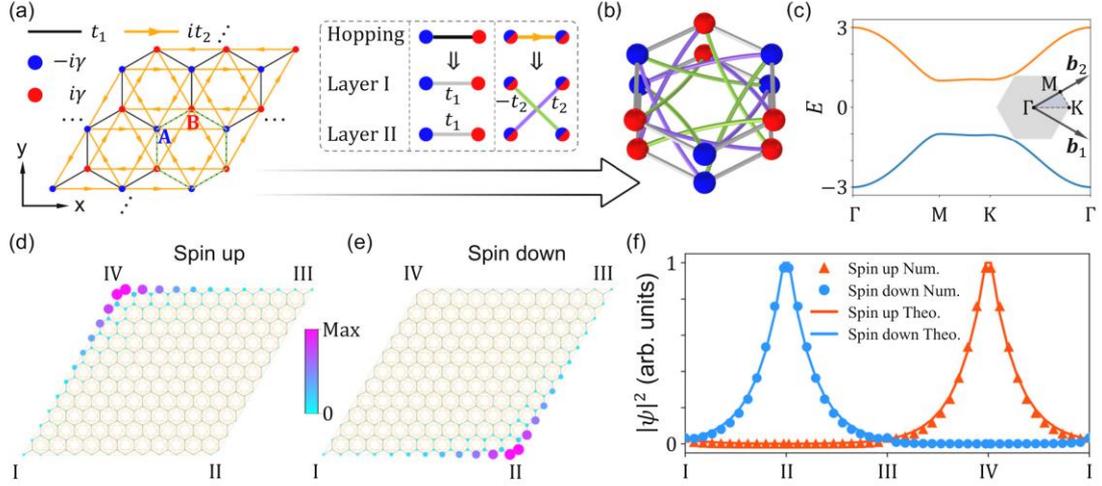

FIG.1 Spin-dependent hybrid skin-topological states in a lattice model. (a) Left panel: Schematic of the Haldane model with loss on sublattice A (blue) and gain on sublattice B (red). Right panel: Scheme of the layer pseudospin implementation, which extends the single layer model to a bilayer one, mimicking the Kane-Mele model. (b) Unit cell of the bilayer lattice model. (c) Bulk band structure of the bilayer lattice model without exerting gain and loss, showing doubly degeneracy for each band. Inset is the first Brillouin zone with reciprocal lattice vectors $\boldsymbol{b}_1$ and $\boldsymbol{b}_2$. (d) and (e) HSEs for the spin-up and -down states, respectively. The spin polarized edge states accumulate separately on opposite corners of a diamond-shaped sample. The size and color of the dot on each site represent the localization strength of the states. (f) Theoretical (curves) and numerical (symbols) results for the spin-dependent hybrid skin-topological states along the closed path I → II → III → IV → I. The parameters here are chosen as $t_1 = 1$, $t_2 = 0.2$, and $\gamma = 0.2$.

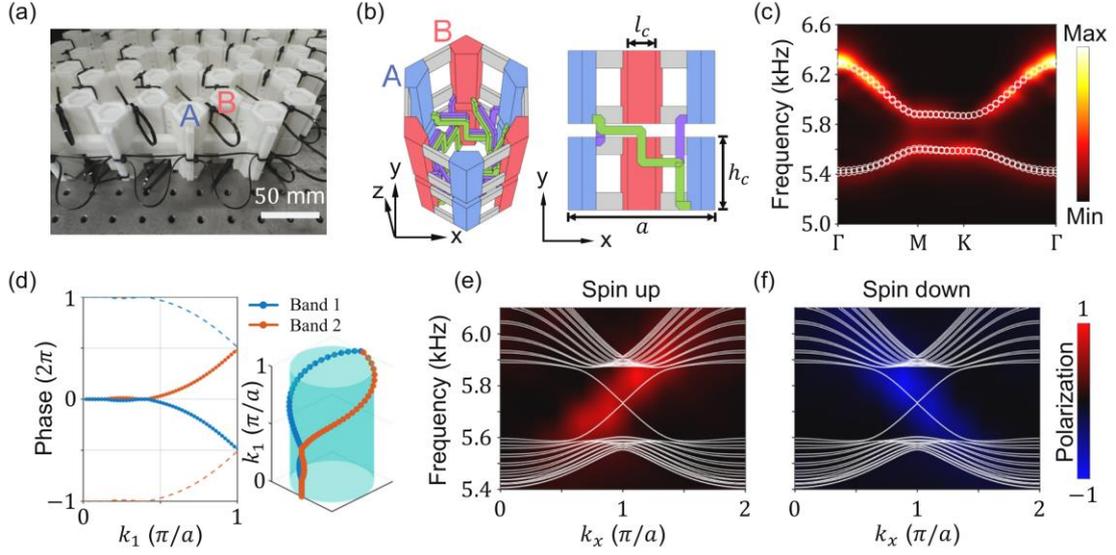

FIG.2 Acoustic implementation of the Kane-Mele model. (a) A photo of the bilayer PC sample. The black cable ties are used to fix the PC sample. (b) Top and front views of the unit cell of the PC (air-filled region). The purple and green waveguide connections act as the positive and negative couplings in the lattice model. (c) Measured bulk dispersions of the PC. The white circles denote four simulated bands of the dipole modes. (d) Wannier center evolutions of the first two bands in (c). Inset shows a mapping to a cylindrical tube. (e) and (f) Measured (colormaps) and simulated (white curves) edge state dispersions. Red and blue colors represent the measured spectra of acoustic pressure fields $p_\pm$ and the spectra are normalized with respect to their maximum values.

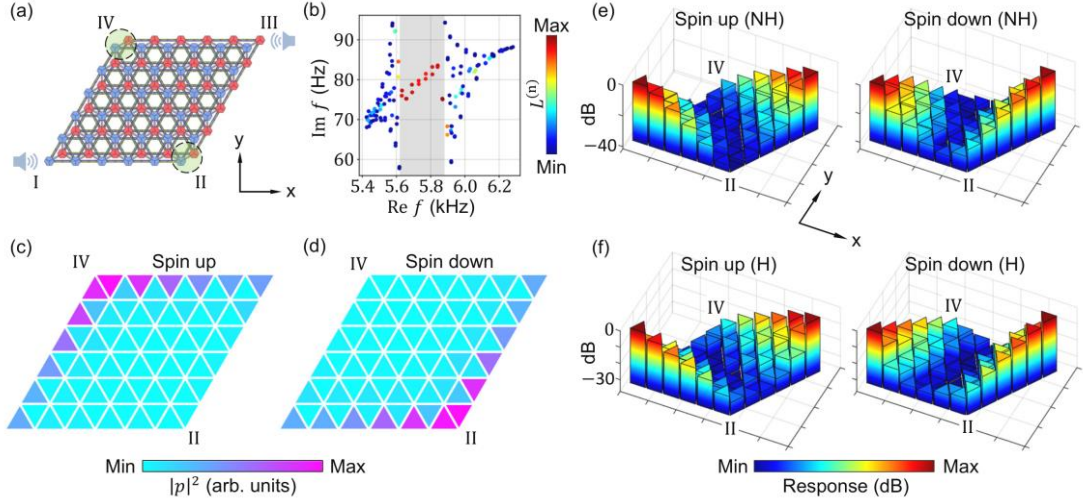

FIG.3 Observation of the SHSE in a non-Hermitian PC. (a) Schematic of the finite PC sample. (b) Simulated complex spectrum of the finite sample. The dot colors denote the localization measure $L^{(n)}$ for the eigenmodes. Gray region labels the bulk gap ranging from 5.65 kHz to 5.85 kHz, where the spin-dependent hybrid skin-topological modes reside. (c) and (d) HSE for the spin up and down states. Colors represent the simulated fields obtained by summing up the squared magnitudes of all the spin polarized eigenmode in the bulk gap. (e) Measured spin-polarized response fields $P_\pm$ in the non-Hermitian PC. It shows that the spin up and down states prefer to accumulate at the corners IV and II, respectively. (f) Same as (e) but measured in a Hermitian PC, which, as a comparison, fails to exhibit inequivalent responses at those corners.

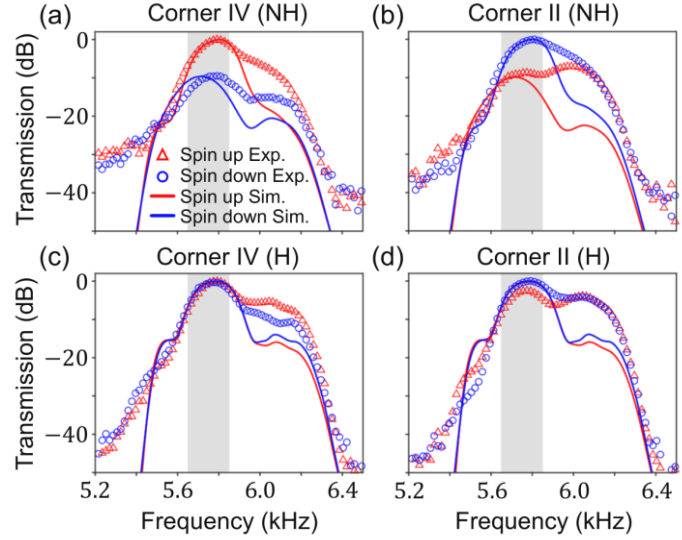

FIG.4 Spin-dependent transmission spectra at the corners II and IV. (a) and (b) Measured (hollow symbols) and simulated (solid curves) transmission spectra gathered respectively from the corners IV and II in the non-Hermitian PC. Differences in the spectra manifest that the spin-up (spin-down) states tend to localize at the corner IV (II). (c) and (d) Same as (a) and (b) but for the Hermitian case. The simulated and experimental results show great agreement in the frequency range of bulk gap (gray region).